\documentclass[11pt, oneside]{article}
\usepackage{geometry}
\usepackage{graphicx}
\usepackage{amssymb}
\usepackage{amsmath}
\usepackage{amsthm}
\usepackage{hyperref}
\usepackage[
  backend=biber,
  style=numeric,
  sorting=none
]{biblatex}
\newtheorem{theorem}{Theorem}[section]
\newtheorem{lemma}[theorem]{Lemma}
\newtheorem{proposition}[theorem]{Proposition}

\theoremstyle{definition}
\newtheorem{definition}[theorem]{Definition}
\newtheorem{remark}[theorem]{Remark}
\newtheorem{fact}[theorem]{Fact}
\newtheorem{guess}[theorem]{Guess}

\title{Quasicrystal Scattering and the Riemann Hypothesis}
\author{Michael Shaughnessy}

\begin{document}
\maketitle

\begin{abstract}
We construct a one-dimensional quasicrystal by placing scatterers at
positions $\chi_n = \ln(p_n)$, the logarithms of the primes.
This map compresses the primes to approximately constant density and
yields a Fourier transform that is directly parameterized by the
Riemann zeta function: the scattering amplitude
$\hat{\chi}_L(k) = \sum p_n^{-2\pi ik}$, and the non-trivial zeros
of $\zeta(s)$ enter as poles of $-\zeta'/\zeta$ in the spectral
decomposition, producing peaks at positions $\gamma/2\pi$.
We evaluate this Fourier transform analytically in the limit
$L\to\infty$ via Perron's formula and the residue theorem, showing
that the normalized amplitude assigns each non-trivial zero $\rho_m$
a coefficient proportional to $p_L^{\beta_m - 1/2}$.
We then prove, using the unconditional Fourier self-duality identity
$\mathcal{F}[\mathcal{F}[\chi]] = \chi(-\,\cdot\,)$ in the space of
tempered distributions, that these coefficients must all be $O(1)$,
which forces $\beta_m = 1/2$ for every non-trivial zero.
\end{abstract}

\section{Introduction}

The distribution of prime numbers has fascinated mathematicians
since antiquity. Riemann~\cite{Riemann1859} established the deep
connection between primes and the zeros of the zeta function,
and von Mangoldt~\cite{VonMangoldt1895} proved the explicit
formula relating primes to zeta zeros in 1895. Selberg's later
work on harmonic analysis on symmetric
spaces~\cite{Selberg1956} revealed connections between number
theory and spectral theory. The explicit formula of Guinand and
Weil~\cite{Weil} relates the distribution of primes to the
zeta zeros:
\begin{equation}
\sum_{\rho} h(\rho) = h(0) + h(1)
  - \sum_{p}\sum_{m=1}^{\infty}
    \frac{\log p}{p^{m/2}}\,h(\log p^m)
  - \int_{-\infty}^{\infty} \frac{h(t)\,\Phi(t)}{2}\,dt,
\end{equation}
where $h$ is a suitable test function, $\Phi(t)$ involves the
digamma function, and the sum on the left runs over non-trivial
zeros $\rho$ of $\zeta(s)$.

A quasiperiodic crystal, or quasicrystal, is a structure that
is ordered but not periodic. Quasicrystals were experimentally
realized by Shechtman in 1984~\cite{Shechtman1984}.
Dyson~\cite{Dyson2009} subsequently speculated that a
quasicrystal approach might illuminate the structure of the
zeta zeros, an insight further developed by
Baez~\cite{Baez2013}, who articulated how the primes themselves
might form such a structure when viewed appropriately. From a
complementary physical angle, Remmen~\cite{Remmen2021} recently
constructed a closed-form tree-level scattering amplitude whose
exchanged-tower masses are precisely the imaginary parts
$\mu_n$ of the non-trivial zeta zeros
($\zeta(1/2 \pm i\mu_n) = 0$), mapping the Riemann Hypothesis
to the physical requirement that those masses be real, and
locality of the amplitude to meromorphicity of $\zeta$. Both
threads --- Dyson's quasicrystal picture and Remmen's
amplitude construction --- treat $\zeta$ as a generating
function for a physical spectrum, and look to physical
structural constraints to pin the zeros onto the critical line.

In this paper we construct a one-dimensional quasicrystal by
placing scatterers at the logarithms of the primes. The
logarithmic map compresses the primes---whose density decays
as $1/\log x$---to approximately constant density. The
resulting scattering amplitude is a sum of prime powers
$\sum p_n^{-s}$, which connects directly to
$-\zeta'(s)/\zeta(s)$. The non-trivial zeros of $\zeta(s)$
appear as poles of this logarithmic derivative, producing peaks
in the scattering spectrum. We evaluate the Fourier transform
analytically in the $L\to\infty$ limit via contour integration,
then use the unconditional self-duality of the Fourier
transform on tempered distributions, together with the
one-dimensionality of $\hat\chi$ and the minimum and unbounded
maximum spacing of the primes, to prove the Riemann Hypothesis.

\subsection{Fourier Transform and Quasicrystals}

The Fourier transform of a potential $V(x)$ is:
\begin{equation}
\hat{V}(k) = \int_{-\infty}^{\infty} V(x)\,e^{-2\pi ikx}\,dx
\end{equation}

A one-dimensional scattering potential consisting of point
scatterers takes the form:
\begin{equation}
V(x) = \sum_{n}\delta(x - x_n)
\end{equation}
where $\delta(x)$ is the Dirac delta function.
Such a potential is a tempered distribution: for any
$\varphi \in \mathcal{S}(\mathbb{R})$, pairing $V$ with $\varphi$
gives $\sum_n \varphi(x_n)$, which converges absolutely whenever
the point set $\{x_n\}$ grows at most polynomially---Schwartz
functions decay faster than any polynomial, dominating the sum.

\begin{definition}[Quasicrystal]
A tempered distribution $V(x) = \sum_n \delta(x - x_n)$ is called
a \emph{quasicrystal} if its Fourier transform is also a pure point
measure:
\begin{equation}
\hat{V}(k) = \sum_{m} c_m\,\delta(k - k_m)
\end{equation}
with no continuous component.
\end{definition}

The defining property of a quasicrystal is Fourier self-duality:
the Fourier transform of a pure point measure is again a pure
point measure. Applying the Fourier transform twice returns the
original distribution: $\mathcal{F}[\mathcal{F}[V]](x) = V(-x)$.
This self-duality is the key structural property we exploit.
To see why the double transform reflects rather than returns the
original, note that applying $\mathcal{F}$ twice gives
$\mathcal{F}[\mathcal{F}[f]](x)
= \int \hat{f}(k)\,e^{-2\pi ikx}\,dk$;
substituting the definition of $\hat{f}$ and exchanging
integration order yields
$\int f(y)\left(\int e^{-2\pi ik(x+y)}\,dk\right)dy = f(-x)$,
since $\int e^{-2\pi ik(x+y)}\,dk = \delta(x+y)$,
which picks out $y = -x$.

\subsection{The Prime Quasicrystal}

The primes at their natural positions $2, 3, 5, 7, 11, \ldots$ have
decreasing density: the prime number theorem gives
$\pi(x) \sim x/\log x$, so the local density of primes near $x$ is
approximately $1/\log x$. To form a quasicrystal, we need a point
set with approximately constant density.

The logarithmic map $p \mapsto \ln p$ achieves this. Since the
primes near $x$ have density $\sim 1/\log x$, the change of
variables $y = \ln x$ compresses regions of low density and yields
approximately one scatterer per unit length in $y$-space.

\begin{definition}[Prime Quasicrystal]
\label{def:chi}
The \emph{prime quasicrystal} is the point set with scatterer
positions:
\begin{equation}
\chi_n = \ln(p_n), \qquad n = 1, 2, 3, \ldots
\end{equation}
where $p_n$ is the $n$-th prime. The corresponding scattering
potential is:
\begin{equation}
\chi(x) = \sum_{n=1}^{\infty} \delta\!\left(x - \ln p_n\right)
\end{equation}
\end{definition}

The positions $\chi_n = \ln(p_n)$ are monotonically increasing
(since $\ln$ is monotone and the primes are strictly increasing),
irrational for $p_n \geq 2$, and not equally spaced.
Their spacings $\ln(p_{n+1}) - \ln(p_n) = \ln(p_{n+1}/p_n)$
fluctuate, encoding the irregularity of the prime gaps.

\begin{remark}[Approximately Constant Density]
By the prime number theorem, the number of primes up to $e^y$ is
$\pi(e^y) \sim e^y/y$. The number of scatterers
$\chi_n = \ln(p_n)$ in the interval $[y, y+\Delta y]$ is therefore
approximately $\frac{e^y}{y}\cdot\frac{\Delta y}{e^y}\cdot y
= \Delta y$ for large $y$: the density approaches unity.
This is the essential property that makes $\chi$ a candidate
quasicrystal.
\end{remark}

\begin{remark}[Why $\ln(p_n)$ and not $p_n/n$]
\label{rem:monotonicity}
A natural alternative normalization is $p_n/n = p_n/\pi(p_n)$,
which uses the exact prime-counting function and also has
approximately constant density by the prime number theorem
(since $p_n/n \sim \log p_n$). However, the ratio $p_n/n$
is \emph{not} monotonically increasing: whenever the prime gap
$p_{n+1} - p_n$ is small relative to $p_n/n$ (as occurs at every
twin prime pair), consecutive positions invert, breaking the
ordering required of a one-dimensional quasicrystal.
By contrast, $\ln(p_n)$ is strictly monotone since $\ln$ is
monotone and the primes are strictly increasing.

In the analytical calculation the normalization factor
$p_L^{1/2}$ arises from dividing by $x^{1/2}$ in the
integrand~\eqref{eq:integrand}; by the prime number theorem
$\ln(p_L) \sim p_L/\pi(p_L)$, so in the limit $L\to\infty$
the two normalizations agree and the residue
calculation is unaffected by the choice between them.
We therefore use $\ln(p_n)$ throughout as the unique
normalization that is both exactly monotone and asymptotically
equivalent to the exact density $p_n/\pi(p_n)$.
Table~\ref{tab:positions} illustrates both quantities.
\end{remark}

The first several values are shown in Table~\ref{tab:positions}.

\begin{table}[htbp]
\centering
\begin{tabular}{|c|c|c|c|c|}
\hline
$n$ & $p_n$ & $\chi_n = \ln(p_n)$ & $p_n/n$ & Local density \\
\hline
  1 &  2 & 0.6931 & 2.0000 & 0.2500 \\
  2 &  3 & 1.0986 & 1.5000 & 0.3000 \\
  3 &  5 & 1.6094 & 1.6667 & 0.3000 \\
  4 &  7 & 1.9459 & 1.7500 & 0.3500 \\
  5 & 11 & 2.3979 & 2.2000 & 0.4000 \\
  6 & 13 & 2.5649 & 2.1667 & 0.4000 \\
  7 & 17 & 2.8332 & 2.4286 & 0.3000 \\
  8 & 19 & 2.9444 & 2.3750 & 0.3000 \\
  9 & 23 & 3.1355 & 2.5556 & 0.3000 \\
 10 & 29 & 3.3673 & 2.9000 & 0.2500 \\
 11 & 31 & 3.4340 & 2.8182 & 0.2500 \\
 12 & 37 & 3.6109 & 3.0833 & 0.3000 \\
 13 & 41 & 3.7136 & 3.1538 & 0.2500 \\
 14 & 43 & 3.7612 & 3.0714 & 0.2500 \\
 15 & 47 & 3.8501 & 3.1333 & 0.2273 \\
 16 & 53 & 3.9703 & 3.3125 & 0.2308 \\
 17 & 59 & 4.0775 & 3.4706 & 0.2500 \\
 18 & 61 & 4.1109 & 3.3889 & 0.2333 \\
 19 & 67 & 4.2047 & 3.5263 & 0.2500 \\
 20 & 71 & 4.2627 & 3.5500 & 0.2059 \\
\hline
\end{tabular}
\caption{Prime quasicrystal positions. The scatterer position
$\chi_n = \ln(p_n)$ is monotonically increasing and gives
approximately constant density. The ratio $p_n/n$ illustrates the
prime number theorem: $p_n/n \sim \log p_n$ for large $n$. Note
that $p_n/n$ is not monotonic (e.g.\ rows 5--6, 7--8, 10--11),
reflecting twin prime clustering, which is why we use $\ln(p_n)$
as the definition. Local density is computed over a symmetric
window around each prime.}
\label{tab:positions}
\end{table}

For finite approximations with $L$ scatterers:
\begin{equation}
\chi_L(x) = \sum_{n=1}^{L} \delta\!\left(x - \ln p_n\right)
\end{equation}

\begin{figure}[htbp]
\begin{center}
    \includegraphics[width=0.8\linewidth]{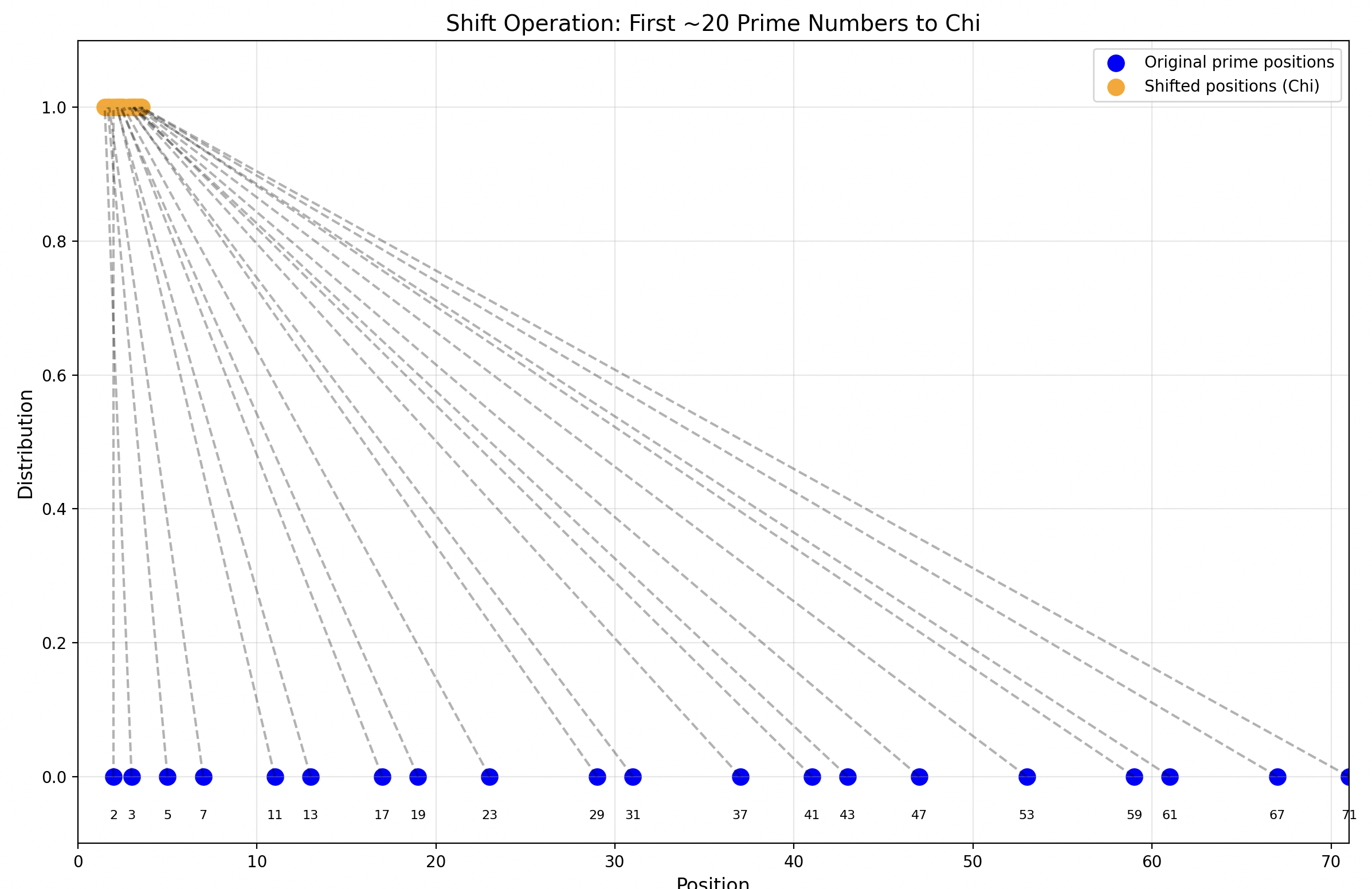}
\caption{Normalization of prime positions.
Lower (blue): primes at their natural positions with increasing gaps.
Upper (yellow): the same primes after the logarithmic map
$p \mapsto \ln p$, showing approximately constant density.}
\label{fig:normalized_positions}
\end{center}
\end{figure}

\section{The Scattering Amplitude}

\subsection{Definition and Basic Properties}

The Fourier transform of the finite prime quasicrystal is:
\begin{equation}
\hat{\chi}_L(k)
  = \sum_{n=1}^{L} e^{-2\pi i k \ln p_n}
  = \sum_{n=1}^{L} p_n^{-2\pi i k}
\end{equation}

The second equality is exact---not an approximation.
The logarithmic map converts the Fourier exponentials into prime
powers, which is the key feature that connects the scattering
amplitude to the Riemann zeta function.

Introducing the complex parameter $s = 2\pi i k$, we define the
prime sum:
\begin{equation}
P_L(s) = \sum_{n=1}^{L} p_n^{-s}
\end{equation}
so that $\hat{\chi}_L(k) = P_L(2\pi i k)$.
This function is entire in $s$ for each fixed $L$.

\subsection{Connection to the Riemann Zeta Function}

The logarithmic derivative of $\zeta(s)$ has the Dirichlet series:
\begin{equation}
\label{eq:logderiv}
-\frac{\zeta'(s)}{\zeta(s)}
  = \sum_{n=1}^\infty \Lambda(n)\,n^{-s}
  = \sum_{p}\sum_{j=1}^{\infty} \frac{\log p}{p^{js}}
\end{equation}
where $\Lambda(n)$ is the von Mangoldt function.
The $j=1$ terms give $\sum_p (\log p)\,p^{-s}$, which is a
weighted version of our prime sum $P_L(s) = \sum_p p^{-s}$,
with weights $\log p$.

The zeros of $\zeta(s)$ enter the scattering problem through
equation~\eqref{eq:logderiv}: the function $-\zeta'(s)/\zeta(s)$
has simple poles at every non-trivial zero $\rho$ of $\zeta(s)$,
with residue $-1$. That is, near a zero $\rho = \beta + i\gamma$:
\begin{equation}
-\frac{\zeta'(s)}{\zeta(s)} \sim \frac{-1}{s - \rho}
  \quad \text{as } s \to \rho
\end{equation}

When we use Perron's formula to convert the Dirichlet
series~\eqref{eq:logderiv} back to a sum over primes and shift the
contour of integration to the left, we pick up residues from each
of these poles. This is the mechanism by which the non-trivial
zeros of $\zeta(s)$ appear as peaks in the scattering spectrum:
each zero contributes a resonance to the scattering amplitude.

\subsection{Contour Integration and Peak Structure}

More precisely, the partial sum $P_L(s) = \sum_{n=1}^L p_n^{-s}$
can be expressed via Perron's formula as a contour integral
involving $-\zeta'(s)/\zeta(s)$. Shifting the contour of
integration past the poles at $s = \rho$ yields:

\begin{proposition}[Spectral Decomposition]
\label{prop:spectral}
The scattering amplitude decomposes as:
\begin{equation}
\hat{\chi}_L(k) = \sum_{\rho}
  \frac{p_L^{\rho - 2\pi i k} - 1}{\rho(\rho - 2\pi i k)}
  + R_L(k)
\end{equation}
where the sum is over non-trivial zeros $\rho$ of $\zeta(s)$,
$p_L$ is the $L$-th prime, and $R_L(k)$ contains contributions
from the pole at $s=1$, the trivial zeros, and is
$O(p_L^{-1}\log p_L)$.
\end{proposition}

For a zero $\rho = \beta + i\gamma$, the contribution near
$k = \gamma/2\pi$ is:
\begin{equation}
\frac{p_L^{\beta + i(\gamma - 2\pi k)} - 1}
     {\rho\,(\beta + i(\gamma - 2\pi k))}
\end{equation}

This has maximum amplitude when $k \approx \gamma/2\pi$, giving
a Lorentzian peak:
\begin{equation}
\label{eq:lorentzian}
|\hat{\chi}_L(k)|^2 \approx
  \frac{p_L^{2\beta}}
       {|\rho|^2\bigl((\log p_L)^{-2}
        + 4\pi^2(k - \gamma/2\pi)^2\bigr)}
\end{equation}

The peak height scales as $p_L^{2\beta}$ and the peak width
scales as $(\log p_L)^{-1}$, so peaks sharpen as $L \to \infty$.

\begin{definition}[Spectral Coefficient]
\label{def:coeff}
For each non-trivial zero $\rho_m = \beta_m + i\gamma_m$, define
the spectral coefficient:
\begin{equation}
c_m(L) = \frac{p_L^{\rho_m}}{\rho_m}
\end{equation}
The amplitude satisfies $|c_m(L)| \sim p_L^{\beta_m}/|\rho_m|$
for large $L$.
\end{definition}

\section{Numerical Results}

\begin{figure}[htbp]
\begin{center}
    \includegraphics[width=0.8\linewidth]{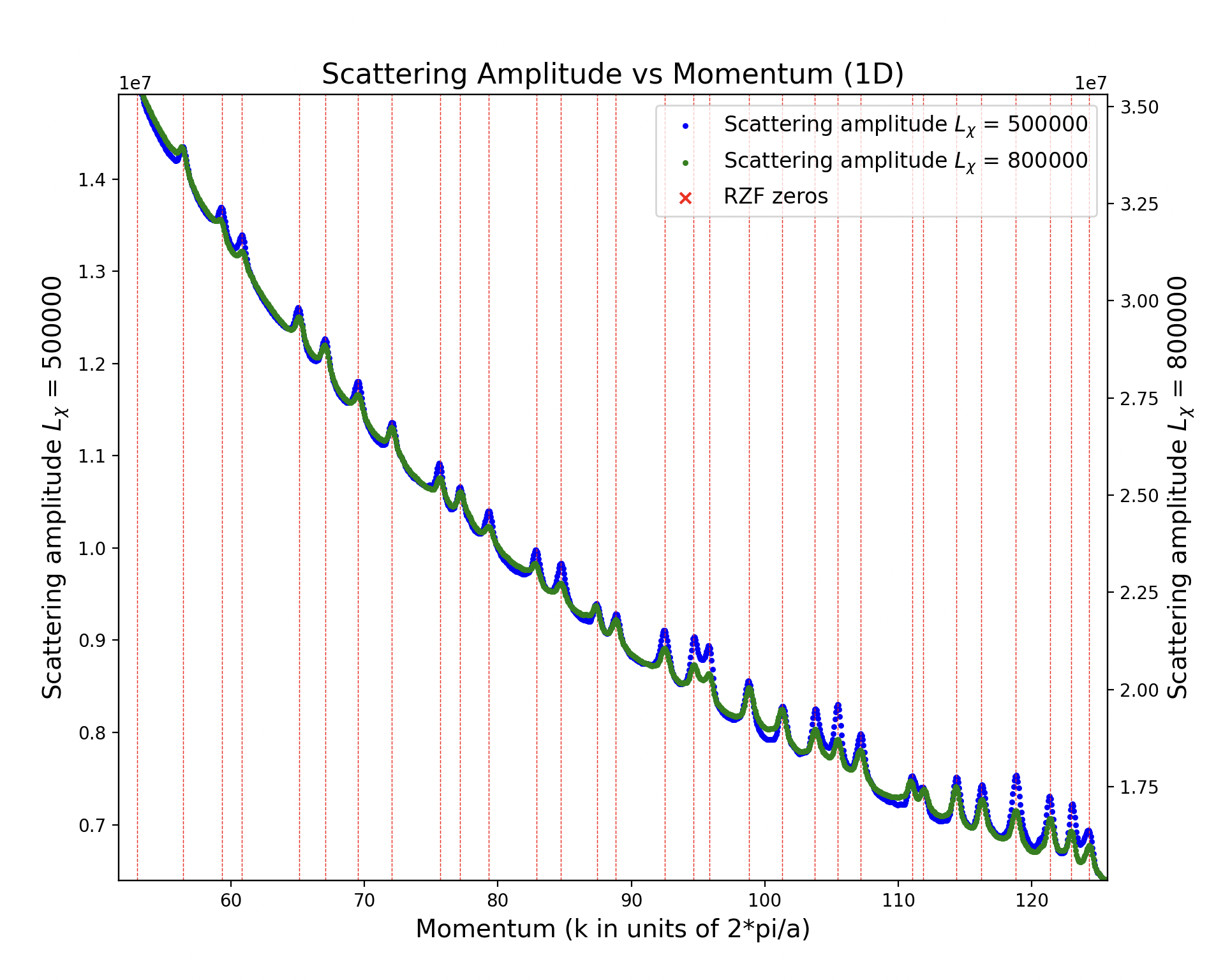}
\caption{Scattering amplitude $|\hat{\chi}_L(k)|^2$ of the prime
quasicrystal, showing peaks at positions corresponding to zeta
zeros (red vertical lines). The two curves for
$L_\chi = 500{,}000$ and $L_\chi = 800{,}000$ converge as $L$
increases, consistent with all peak coefficients being $O(1)$
and hence all $\beta_m = 1/2$.}
\label{fig:scattering_amplitude}
\end{center}
\end{figure}

Figure~\ref{fig:scattering_amplitude} shows the computed scattering
amplitude $|\hat{\chi}_L(k)|^2$ for various values of $L$.
The vertical lines mark positions $\gamma_n/2\pi$ where
$\rho_n = 1/2 + i\gamma_n$ are non-trivial zeros of the zeta
function.

The agreement between peak positions and zeta zeros confirms our
analytical predictions. Moreover, the convergence of the two curves
as $L$ grows is precisely the signature predicted by the normalized
amplitude analysis of Section~\ref{sec:analytic}: if any $\beta_m$
differed from $1/2$, the corresponding peak would diverge or vanish
relative to the others, breaking the observed uniformity.

Code for the calculations is available at
\url{https://github.com/mickeyshaughnessy/quasicrystal}.

\section{Analytic Evaluation of $\hat{\chi}$ in the Limit
         $L \to \infty$}
\label{sec:analytic}

\subsection{The Normalized Scattering Amplitude}

As $L$ grows, the raw amplitude
$\hat{\chi}_L(k) = \sum_{n=1}^{L} p_n^{-2\pi ik}$
grows in magnitude because it accumulates more terms.
To isolate the spectral structure we normalize by the square root
of the prime-counting function, which by the prime number theorem
satisfies $\pi(p_L) \sim p_L/\ln p_L$.
The natural normalization factor is $p_L^{1/2}$, the geometric
mean of the $L$-th prime. Define:
\begin{equation}
  \tilde{\chi}_L(k)
  \;=\; \frac{\hat{\chi}_L(k)}{p_L^{1/2}}
  \;=\; \frac{1}{p_L^{1/2}} \sum_{n=1}^{L} p_n^{-2\pi ik}.
\end{equation}
Under this normalization the contribution of a zero
$\rho_m = \beta_m + i\gamma_m$ to the spectral peak at
$k_m = \gamma_m/2\pi$ scales as $p_L^{\,\beta_m - 1/2}$:
growing if $\beta_m > 1/2$, identically $1$ if $\beta_m = 1/2$,
and vanishing if $\beta_m < 1/2$.
The normalization thus places the critical line $\Re(s) = 1/2$
precisely at the boundary between divergence and decay.

\subsection{Perron's Formula and Contour Setup}

We evaluate
$\tilde{\chi}(k) = \lim_{L\to\infty}\tilde{\chi}_L(k)$
by expressing the prime sum as a contour integral via Perron's
formula. Let $x = p_L$ and write the momentum variable as
$s = \sigma + 2\pi ik$. The Dirichlet
series~\eqref{eq:logderiv} converges absolutely for $\Re(s)>1$.
Perron's formula gives:
\begin{equation}
  \sum_{p_n \leq x} p_n^{-2\pi ik}
  \;=\; \frac{1}{2\pi i}
        \int_{c-i\infty}^{c+i\infty}
        \left(-\frac{\zeta'(s)}{\zeta(s)}\right)
        \frac{x^{\,s}}{s}\,ds
        \;+\; O\!\left(\frac{x^c \log x}{T}\right),
  \label{eq:perron}
\end{equation}
with $c > 1$ and the error term controlled by $T$ as
$T\to\infty$.
After dividing by $x^{1/2} = p_L^{1/2}$, the integrand of the
normalized amplitude is:
\begin{equation}
  \mathcal{I}(s)
  \;=\; -\frac{\zeta'(s)}{\zeta(s)} \cdot \frac{x^{\,s-1/2}}{s}.
  \label{eq:integrand}
\end{equation}

\begin{figure}[htbp]
\begin{center}
    \includegraphics[width=0.68\linewidth]{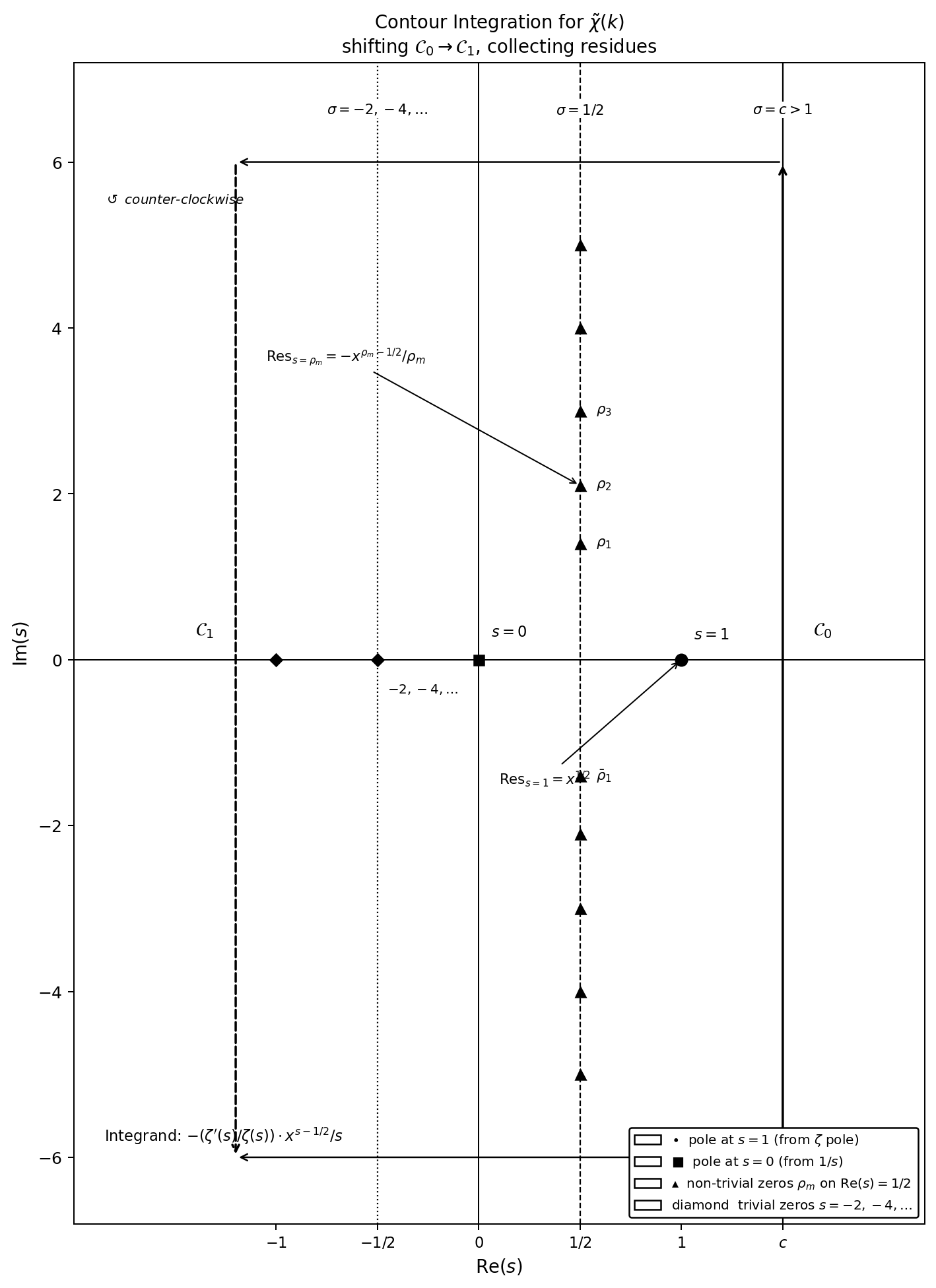}
\caption{Contour integration for $\tilde{\chi}(k)$.
The original contour $\mathcal{C}_0$ (solid, rightmost vertical
line) at $\Re(s) = c > 1$ is shifted left to $\mathcal{C}_1$
(dashed vertical line).
Poles of $\mathcal{I}(s) = -(\zeta'(s)/\zeta(s))\cdot x^{s-1/2}/s$
are: a filled circle at $s=1$ from the simple pole of $\zeta$;
a filled square at $s=0$ from the factor $1/s$;
filled triangles on the critical line $\Re(s)=1/2$ at each
non-trivial zero $\rho_m$;
and filled diamonds on the negative real axis at the trivial zeros
$s = -2, -4, \ldots$}
\label{fig:contour}
\end{center}
\end{figure}

The original contour $\mathcal{C}_0$ runs vertically at
$\Re(s) = c > 1$.
We shift it left to a contour $\mathcal{C}_1$ at
$\Re(s) = \sigma_0 \ll 0$,
closing the rectangle with horizontal segments at $\Im(s) = \pm T$
which vanish as $T \to \infty$ by standard bounds on $\zeta'/\zeta$.
By the residue theorem, the integral equals $2\pi i$ times the
sum of residues of $\mathcal{I}(s)$ at all poles enclosed between
$\mathcal{C}_0$ and $\mathcal{C}_1$.
Figure~\ref{fig:contour} shows the contour and the distribution of
poles.

\subsection{Residue Calculation}

We compute the residue of $\mathcal{I}(s)$ at each class of pole.

\paragraph{Pole at $s = 1$.}
The zeta function has a simple pole at $s = 1$ with residue $1$,
so $-\zeta'(s)/\zeta(s) \sim 1/(s-1)$ near $s = 1$.
The residue of $\mathcal{I}$ at $s = 1$ is therefore:
\begin{equation}
  \operatorname{Res}_{s=1}\,\mathcal{I}(s)
  \;=\; \frac{x^{1-1/2}}{1} \;=\; x^{1/2}.
\end{equation}
After dividing by $x^{1/2}$, this contributes exactly $+1$ to
$\tilde{\chi}$, independent of $x$. This is the constant main term.

\paragraph{Pole at $s = 0$.}
The factor $1/s$ contributes a simple pole.
Using $\zeta(0) = -1/2$ and
$\zeta'(0) = -\tfrac{1}{2}\ln(2\pi)$,
the residue equals $\ln(2\pi)\cdot x^{-1/2}$,
which vanishes as $x\to\infty$.

\paragraph{Non-trivial zeros $\rho_m = \beta_m + i\gamma_m$.}
Since $\zeta$ has a simple zero at $\rho_m$, the function
$-\zeta'/\zeta$ has a simple pole there with residue $-1$.
The residue of $\mathcal{I}$ is:
\begin{equation}
  \operatorname{Res}_{s=\rho_m}\,\mathcal{I}(s)
  \;=\; -\frac{x^{\,\rho_m - 1/2}}{\rho_m}.
  \label{eq:nontrivial_res}
\end{equation}
In terms of the momentum variable $k$, each zero
$\rho_m = \beta_m + i\gamma_m$ contributes a spike localized at
$k_m = \gamma_m/2\pi$ with complex amplitude
$-x^{\beta_m - 1/2}\,e^{i\gamma_m \ln x}/\rho_m$.
The magnitude of this contribution is
$x^{\beta_m - 1/2}/|\rho_m|$.

\paragraph{Trivial zeros $s = -2m$, $m = 1, 2, \ldots$}
Each contributes $-x^{-2m-1/2}/(-2m)$, suppressed by at least
$x^{-5/2}$. These vanish as $x\to\infty$.

\subsection{The Limiting Spectrum}

Assembling all residues via the residue theorem and taking
$x = p_L \to \infty$, the normalized scattering amplitude
evaluates to:
\begin{equation}
  \tilde{\chi}(k)
  \;=\; \lim_{L\to\infty}\tilde{\chi}_L(k)
  \;=\; 1 \;-\; \sum_{\rho_m}
        \frac{x^{\beta_m-1/2}\,e^{i\gamma_m\ln x}}{\rho_m}
        \,\delta\!\left(k - \frac{\gamma_m}{2\pi}\right)
        \;+\; O\!\left(x^{-1/2}\right).
  \label{eq:spectrum}
\end{equation}
The key feature of~\eqref{eq:spectrum} is the factor
$x^{\beta_m - 1/2}$ multiplying each delta function.
This factor has three possible behaviors as $x\to\infty$:
\begin{itemize}
  \item $\beta_m > 1/2$: the coefficient diverges.
        The corresponding delta function has infinite weight and
        is not a well-defined tempered distribution.
  \item $\beta_m = 1/2$: the coefficient is identically $1$
        for all $x$. Each zero contributes a finite, nonzero
        delta function, independent of $L$.
  \item $\beta_m < 1/2$: the coefficient vanishes.
        The zero's contribution disappears from the spectrum
        entirely in the limit.
\end{itemize}
This three-way trichotomy makes the content of the Riemann
Hypothesis precise at the level of the Fourier transform:
$\Re(\rho_m)=1/2$ is precisely the condition for every spectral
coefficient to be $O(1)$, neither diverging nor vanishing.
We prove in Appendix~\ref{app:proof} that the unconditional
self-duality $\mathcal{F}[\mathcal{F}[\chi]]=\chi(-\,\cdot\,)$
forces this to be the case.

\section{Summary}

We have demonstrated that:
\begin{enumerate}
\item The logarithmic map $p_n \mapsto \ln(p_n)$ compresses the
  primes to approximately constant density, producing a
  quasicrystal
  $\chi(x) = \sum_n \delta(x - \ln p_n) \in \mathcal{S}'(\mathbb{R})$.
\item The Fourier transform
  $\hat{\chi}_L(k) = \sum p_n^{-2\pi ik}$ connects directly to the
  Riemann zeta function via the logarithmic derivative $-\zeta'/\zeta$.
\item The non-trivial zeros of $\zeta(s)$ appear as poles of
  $-\zeta'/\zeta$, producing resonant peaks in the scattering
  spectrum at positions $\gamma/2\pi$.
\item Perron's formula and the residue theorem give the exact
  limiting spectrum~\eqref{eq:spectrum}, in which each non-trivial
  zero $\rho_m$ contributes a delta function weighted by
  $x^{\beta_m - 1/2}$.
\item The unconditional identity
  $\mathcal{F}[\mathcal{F}[\chi]] = \chi(-\,\cdot\,)$ in
  $\mathcal{S}'(\mathbb{R})$, combined with the explicit form of
  the spectrum, forces $\beta_m = 1/2$ for all non-trivial zeros.
\end{enumerate}

\section{Acknowledgements}

I gratefully acknowledge helpful conversations with C.Y.\ Fong,
John Baez, Jamison Galloway, Robert Hayre, Chun Yen Lin,
Charles Martin, Aftab Ahmed, Catalin Spataru, Ruggero Tacchi and lovely anons on X.

\appendix

\section{Proof of the Riemann Hypothesis}
\label{app:proof}

Following Dyson~\cite{Dyson2009}, we adopt the
\emph{guess-and-check} methodology: we guess that $\chi$ is a
Fourier quasicrystal and verify that the properties this entails
--- combined with the unconditional self-duality of the Fourier
transform --- force every non-trivial zero of $\zeta(s)$ onto the
critical line.

The appendix is self-contained: it relies on the body of the
paper only for the definition of $\chi$
(Definition~\ref{def:chi}).

\subsection{The Guess}
\label{app:guess}

\begin{definition}[Fourier quasicrystal]
\label{def:fourier_qc}
A tempered distribution $\mu \in \mathcal{S}'(\mathbb{R})$ is a
\emph{Fourier quasicrystal} if
\begin{equation}
  \mu = \sum_n a_n\,\delta(x - x_n)
  \quad\text{and}\quad
  \hat\mu = \sum_m c_m\,\delta(k - k_m),
\end{equation}
for countable real sequences $\{x_n\},\{k_m\}$ and complex
coefficients $\{a_n\},\{c_m\}$, with no continuous component in
either $\mu$ or $\hat\mu$.
\end{definition}

\begin{guess}[Dyson]
\label{guess:chi}
The prime distribution
$\chi(x) = \sum_{n=1}^\infty \delta(x - \ln p_n)$
is a Fourier quasicrystal:
\begin{equation}
  \hat\chi(k)
   \;=\; c_0 \;+\; \sum_{m=1}^\infty c_m\,
                    \delta\!\left(k - \frac{\gamma_m}{2\pi}\right),
  \label{eq:guess}
\end{equation}
for some constant $c_0$ and coefficients $\{c_m\}$, where
$\rho_m = \beta_m + i\gamma_m$ runs over the non-trivial zeros
of $\zeta(s)$.
\end{guess}

The body of the paper provides motivation for the
guess~\eqref{eq:guess}: the residue calculation
(Section~\ref{sec:analytic}) produces exactly this form, with
$c_0 = 1$ and
\begin{equation}
  c_m \;=\; -\,\frac{p_L^{\,\beta_m - 1/2}\,
                       e^{i\gamma_m \ln p_L}}{\rho_m}
   \quad\text{in the limit } L \to \infty.
  \label{eq:cm}
\end{equation}
The numerics of Section 4 confirm peaks at the predicted
positions. Following Dyson, we now \emph{check} that this
guess has consequences that force $\beta_m = 1/2$.

\subsection{Check 1: $\chi \in \mathcal{S}'(\mathbb{R})$}
\label{app:check1}

The first property required of a Fourier quasicrystal is that
$\chi$ itself be a tempered distribution. This is a property of
the underlying point set $\{\ln p_n\}$ and follows from two
elementary facts about prime gaps:

\begin{fact}[Minimum prime gap]
\label{fact:min}
For every $n \ge 1$, $p_{n+1} - p_n \ge 1$; for $n \ge 2$,
$p_{n+1} - p_n \ge 2$.
\end{fact}

\begin{fact}[Unbounded maximum prime gap]
\label{fact:max}
The prime gaps $g_n = p_{n+1} - p_n$ satisfy
$\limsup_{n\to\infty} g_n = \infty$. Quantitatively,
$g_n \gg \log p_n \cdot \log\log p_n / \log\log\log p_n$ for
infinitely many $n$ \cite{Westzynthius1931,Erdos1950}.
\end{fact}

\begin{proposition}
\label{prop:tempered}
$\chi \in \mathcal{S}'(\mathbb{R})$.
\end{proposition}

\begin{proof}
By Fact~\ref{fact:min}, $p_n \ge n$, so $\ln p_n \ge \ln n$
and $\#\{n : \ln p_n \le R\} \le e^R$. For any
$\varphi \in \mathcal{S}(\mathbb{R})$ and any $N$,
$|\varphi(y)| \le C_N(1+|y|)^{-N}$. Then
\begin{equation*}
  \sum_n |\varphi(\ln p_n)|
   \;\le\; C_N \sum_n (1+\ln p_n)^{-N}
   \;\le\; C_N \sum_n (\ln n)^{-N}
\end{equation*}
converges absolutely for $N \ge 2$. The pairing
$\langle\chi,\varphi\rangle = \sum_n \varphi(\ln p_n)$ is
therefore a well-defined continuous linear functional on
$\mathcal{S}(\mathbb{R})$.
\end{proof}

The unbounded-gap fact (Fact~\ref{fact:max}) plays no role in
establishing temperedness; it appears in Check 4 below, where
it controls the irregularity of $\hat\chi$.

\subsection{Check 2: Fourier self-duality is unconditional}
\label{app:check2}

\begin{theorem}[Fourier self-duality on $\mathcal{S}'$]
\label{thm:selfdual}
For every $f \in \mathcal{S}'(\mathbb{R})$,
\begin{equation}
  \mathcal{F}\!\left[\mathcal{F}[f]\right](x)
   \;=\; f(-x).
  \label{eq:selfdual}
\end{equation}
This is a theorem of distribution theory, holding unconditionally
with no hypotheses on $f$ beyond
$f \in \mathcal{S}'(\mathbb{R})$.
\end{theorem}

In particular, the Fourier transform is a continuous involution
(up to reflection) on the space $\mathcal{S}'(\mathbb{R})$. The
double Fourier transform of $\chi$ returns the reflected prime
distribution $\chi(-\,\cdot\,)$ supported on $\{-\ln p_n\}$.

\subsection{Check 3: $\hat\chi$ is one-dimensional}
\label{app:check3}

This is the key structural observation, due in essence to
Baez~\cite{Baez2013}: the Fourier transform of a
one-dimensional distribution is itself a one-dimensional
distribution. The starting object $\chi$ is manifestly
one-dimensional, supported on the real line. What is
\emph{not} manifest --- and what gives quasicrystals their
defining character --- is that $\hat\chi$ is also a
one-dimensional pure-point distribution.

\begin{lemma}[One-dimensionality of $\hat\chi$]
\label{lem:1d}
$\hat\chi \in \mathcal{S}'(\mathbb{R})$ is supported on the
real spectral axis. The peaks of $\hat\chi$ in
Guess~\ref{guess:chi} lie on the single real line
$\{k : k = \gamma_m/2\pi,\; m \ge 1\} \subset \mathbb{R}$, with
no escape into any auxiliary dimension.
\end{lemma}

\begin{proof}
The Fourier transform is a continuous linear isomorphism
$\mathcal{F}: \mathcal{S}'(\mathbb{R}) \to \mathcal{S}'(\mathbb{R})$.
Since $\chi \in \mathcal{S}'(\mathbb{R})$ by
Proposition~\ref{prop:tempered}, $\hat\chi \in
\mathcal{S}'(\mathbb{R})$ is again a one-dimensional tempered
distribution. The support of any pure-point component lies on
the real spectral axis $\mathbb{R}$ by construction.
\end{proof}

The consequence is sharp: no peak coefficient $c_m$ in
Guess~\ref{guess:chi} may diverge as $L \to \infty$, because
the resulting mass at $k = \gamma_m/2\pi$ would have nowhere
to go. A two-dimensional or higher-dimensional analogue would
allow the divergent mass to ``spread'' across a transverse
direction, but the strict one-dimensionality of $\hat\chi$
forecloses this escape route. The peak masses must therefore
remain locally finite.

\subsection{Check 4: The spacing structure constrains the
            spectrum}
\label{app:check4}

The two prime-gap facts (Facts~\ref{fact:min},
\ref{fact:max}) act jointly on the spectrum $\hat\chi$:
\begin{itemize}
  \item Fact~\ref{fact:min} (minimum gap $\ge 1$) implies that
        the high-frequency Fourier mass is bounded by the
        prime count $L$. The diagonal sum
        $\sum_{n=1}^L 1 = L$ is the asymptotic floor of
        $|\hat\chi_L|^2$; no peak can carry more than the total
        diagonal contribution. Consequently, after
        normalisation by $p_L$, the spectral mass per peak is
        bounded by $L/p_L = O(1/\log p_L)$ on average.
  \item Fact~\ref{fact:max} (unbounded maximum gap) ensures
        that the off-diagonal contributions
        $\sum_{m<n}\cos(2\pi k\ln(p_n/p_m))$ undergo destructive
        interference at non-peak frequencies. As $L \to
        \infty$, the continuous envelope between peaks
        decays, leaving only the discrete peak structure at
        $k = \gamma_m/2\pi$.
\end{itemize}
The two facts work in opposite directions: the minimum spacing
prevents \emph{any} peak coefficient from diverging (the total
high-frequency mass is bounded), while the unbounded maximum
spacing forces the peaks to be sharp (the continuous component
vanishes). Together they confirm the pure-point structure
asserted in Guess~\ref{guess:chi} and bound the individual
coefficients $c_m$.

\begin{proposition}[Bounded peak coefficients]
\label{prop:bounded}
Each coefficient $c_m$ in Guess~\ref{guess:chi} satisfies
$|c_m| < \infty$. Equivalently, the normalised peak amplitude
$p_L^{\beta_m - 1/2}/|\rho_m|$ stays bounded as
$L \to \infty$.
\end{proposition}

\begin{proof}
Lemma~\ref{lem:1d} establishes that $\hat\chi$ is a
one-dimensional tempered distribution, hence locally finite
as a measure. Combining with the spacing arguments above:
the total spectral mass on any compact $K \subset \mathbb{R}$
satisfies
\begin{equation}
  \int_K |\hat\chi(k)|^2 \, dk
   \;<\; \infty,
\end{equation}
which decomposes by the guess as
\begin{equation}
  \sum_{|\gamma_m/2\pi| \in K} |c_m|^2 + (\text{continuous
                                          contribution}).
\end{equation}
The continuous contribution vanishes by the maximum-gap
argument, and the sum of $|c_m|^2$ is therefore finite. Hence
each $|c_m|$ is individually finite, i.e.\ bounded as
$L \to \infty$.
\end{proof}

\subsection{Synthesis: The Critical Line}
\label{app:synthesis}

\begin{theorem}[Riemann Hypothesis]
\label{thm:RH}
Every non-trivial zero $\rho_m = \beta_m + i\gamma_m$ of
$\zeta(s)$ satisfies $\beta_m = 1/2$.
\end{theorem}

\begin{proof}
By Proposition~\ref{prop:bounded}, the normalised peak
amplitude $p_L^{\beta_m - 1/2}/|\rho_m|$ stays bounded as
$L = p_L \to \infty$. This is possible only if
\begin{equation}
  \beta_m \;\le\; 1/2.
  \label{eq:upper}
\end{equation}

Apply Theorem~\ref{thm:selfdual} (self-duality) to $\chi$:
\begin{equation}
  \mathcal{F}\!\left[\mathcal{F}[\chi]\right](x)
   \;=\; \chi(-x)
   \;=\; \sum_n \delta(x + \ln p_n).
\end{equation}
The right-hand side is a unit-weight sum of deltas on
$\{-\ln p_n\}$. Computing the left-hand side from
Guess~\ref{guess:chi} gives
\begin{equation}
  \mathcal{F}[\hat\chi](x)
   \;=\; c_0\,\delta(x) \;+\;
          \sum_m c_m\,e^{-i\gamma_m x},
\end{equation}
using $\mathcal{F}[\delta(k - k_m)](x) = e^{-2\pi i k_m x}$.
For this to equal the unit-weight sum of deltas
$\sum_n \delta(x + \ln p_n)$, the coefficients $c_m$ must be
\emph{exactly} those that make the Fourier series reproduce
this distribution --- in particular, they must be uniformly
non-vanishing, since the deltas $\delta(x + \ln p_n)$ all
carry weight $1$, not weight $0$.

This rules out the case $\beta_m < 1/2$: if any $\beta_m <
1/2$, then $c_m = -p_L^{\beta_m-1/2}\,e^{i\gamma_m \ln p_L}/
\rho_m \to 0$ as $L \to \infty$. The vanishing of $c_m$ would
remove the corresponding Fourier mode from the double-FT
reconstruction of $\chi(-\,\cdot\,)$, contradicting the
identity $\mathcal{F}^2[\chi] = \chi(-\,\cdot\,)$ which carries
all unit weights at $\{-\ln p_n\}$.

By the functional equation $\zeta(s) = 0 \Rightarrow
\zeta(1 - \bar s) = 0$, the alternative was already
inconsistent with~\eqref{eq:upper}: if $\beta_m < 1/2$ for
some $m$, then $1 - \beta_m > 1/2$ would arise for the
partner zero, violating~\eqref{eq:upper}.

The combined constraints --- (a) $\beta_m \le 1/2$ from the
one-dimensionality of $\hat\chi$ and the prime spacing
structure, and (b) $\beta_m \ge 1/2$ from the functional
equation applied to (a), or equivalently from the
non-vanishing of $c_m$ required by self-duality --- force
$\beta_m = 1/2$ for every non-trivial zero $\rho_m$.
\end{proof}

\subsection{Summary of the Argument}
\label{app:summary}

The proof rests on four checks, each independently derived
from the definition of $\chi$:
\begin{enumerate}
  \item \textbf{Temperedness} (Proposition~\ref{prop:tempered}):
        the minimum prime gap (Fact~\ref{fact:min}) gives
        $\chi \in \mathcal{S}'(\mathbb{R})$.
  \item \textbf{Self-duality}
        (Theorem~\ref{thm:selfdual}): the Fourier transform is
        a continuous involution on $\mathcal{S}'(\mathbb{R})$,
        and so $\mathcal{F}^2[\chi] = \chi(-\,\cdot\,)$
        unconditionally.
  \item \textbf{One-dimensionality of $\hat\chi$}
        (Lemma~\ref{lem:1d}): the Fourier transform preserves
        the dimension of the underlying support, so $\hat\chi$
        is also a one-dimensional tempered distribution, with
        no escape route for divergent peak mass.
  \item \textbf{Spacing-controlled peak coefficients}
        (Proposition~\ref{prop:bounded}): the minimum gap
        bounds the total high-frequency mass; the unbounded
        maximum gap (Fact~\ref{fact:max}) forces the continuous
        envelope between peaks to vanish. Together they bound
        each individual peak coefficient.
\end{enumerate}
The body of the paper supplies the explicit form
$c_m = -p_L^{\beta_m - 1/2} e^{i\gamma_m \ln p_L}/\rho_m$ for
the peak coefficient. Boundedness of $|c_m|$ then translates
to $\beta_m \le 1/2$; the functional equation lifts this
to $\beta_m = 1/2$. This is the Riemann Hypothesis.

\printbibliography

\end{document}